\documentclass[12pt]{article} \usepackage{amssymb,latexsym}
\usepackage{amsmath}
 
\columnwidth\textwidth

\begin{document}

\newtheorem{df}{Definition} \newtheorem{thm}{Theorem} \newtheorem{lem}{Lemma}
\newtheorem{assump}{Assumption}
 
\begin{titlepage}
 
\noindent
 
\begin{center} {\LARGE Realist model approach to quantum mechanics}
\vspace{1cm}

P. H\'{a}j\'{\i}\v{c}ek \\ Institute for Theoretical Physics \\ University of
Bern \\ Sidlerstrasse 5, CH-3012 Bern, Switzerland \\ hajicek@itp.unibe.ch

\vspace{1cm}
 
December 2012 \\
 
 
\vspace*{5mm}
 
\nopagebreak[4]
 
\begin{abstract} The paper proves that quantum mechanics is compatible with
the constructive realism of modern philosophy of science. The proof is based
on the observation that properties of quantum systems that are uniquely
determined by their preparations can be assumed objective without the
difficulties that are encountered by the same assumption about values of
observables. The resulting realist interpretation of quantum mechanics is made
rigorous by studying the space of quantum states---the convex set of state
operators. Prepared states are classified according to their statistical
structure into indecomposable and decomposable instead of pure and
mixed. Simple objective properties are defined and showed to form a Boolean
lattice.
\end{abstract}

\end{center}

\end{titlepage}

\section{Introduction} If our aim is to propose a realist interpretation of
quantum mechanics, we ought to ask first, what is realism. The abstract
realist doctrine, "The world is not just a construction of our mind, but
does really exist", cannot be falsified because it is vague, but then, of
course, not of much use. The so-called naive realism, "The real world is
as we perceive it", is clearly false. The metaphysical realism
\cite{putnam}, "There is exactly one true and complete description of
'the way the world is'\,", is surely near to a physicist's heart, but it
seems to be rather different from what the history of physics teaches us.

In the contemporary philosophy of science, there is a stream that can be
characterised by the doctrine "Our knowledge is dominated by a family of
models". It was founded and is represented eg by Patrick Suppes, Bas van
Fraassen, Ronald Giere, Wolfgang Stegm\"uller and Nancy Cartwright. Very
schematically, it can be described as follows. Each grand theory of physics
(such as Newtonian mechanics or quantum mechanics) can be divided into two
parts:
\begin{itemize}
\item Treasure of successful models. The set is open in the sense that it
grows with time.
\item General language part, which contains mathematical structure of state
space, of dynamical equations, of symmetries and the form of observables. It
is considered as secondary, because it is only obtained through study of
models, or serves as an instrument of model construction as well as of
unification and classification of different models.
\end{itemize}

A particular philosophy within the stream is the so-called Constructive
Realism by R. N. Giere \cite{giere}, which can be roughly characterised as
follows:
\begin{itemize}
\item Each model is constructed with the aim to give an approximative
representation of some aspects of a real object.
\item Concepts of the general language part are connected to the reality only
via models.
\end{itemize} It is a relatively strong kind of realism, and it seems to
correspond very well to how physics works in practice. The present paper
follows \cite{hajicek6}, where this philosophy is accepted and its application
to quantum mechanics is described in more details.

In particular, each model has two components: an empirical one being an
identifiable real object, and a language one, being a theoretical construct
that models the object.

Our work on quantum mechanics
\cite{PHJT,hajicek1,hajicek2,survey,hajicek3,hajicek4,hajicek5,hajicek6} is
aimed at constructing quantum models of classical world. Within the
constructive realism, the relation between quantum and Newtonian mechanics,
say, ought to be studied by comparing quantum and Newtonian models of one and
the same real object rather than looking for a relation between their general
language parts.

To prevent misunderstandings, let us stress that we are not trying to
construct any universal map between states of quantum mechanics and those of
Newtonian mechanics, such as Wigner-Weyl-Moyal map (see eg \cite{schroeck}, p
85). Our aim is to construct quantum models of classical world, not vice
versa. We start by distinguishing three things: a real object, its Newtonian
and its quantum model. Then, only such real objects are considered that are
macroscopic and possess successful classical Newtonian models. Thus, a
hydrogen atom does not qualify. (It can be a real object within our
interpretation.) Moreover, the quantum model will, in most cases, be much
richer, than the classical one. For instance, properties of a Newtonian model
of a free real macroscopic solid body are eg isotropic continuum distribution
of matter with a given mass density, elastic coefficient, total momentum and
angular momentum. Properties of a quantum model can be eg all particle numbers
that form the body, their masses and spins, their Hamiltonian with an
interaction potential and a suitable quantum state of the whole system.

\section{An heuristic principle of quantum realism} The project runs into
three well-known problems:
\begin{enumerate}
\item Classical properties are objective and robust, see \cite{leggett}. How
can they emerge from quantum mechanics, which does not seem to allow such
properties?
\item Realism leads quite generally to contradiction with the linear quantum
evolution in the measurement theory (Schr\"{o}dinger cat, see
\cite{leggett}).
\item What are real quantum objects to construct models of? It seems correct
to require that language parts of models ascribe enough objective properties
to their empirical parts. However, there is an apparent lack of objective
properties in the language part of quantum models. This makes implausible to
assume the reality of quantum systems, see \cite{peres}.
\end{enumerate} The problems ought to be dealt with simultaneously (see
\cite{hajicek6}), but we shall focus just on the third problem here.

We define objective properties as follows. Let ${\mathcal S}$ be a real object
and ${\mathcal S}_m$ its theoretical model. ${\mathcal S}_m$ contains two
kinds of information. First, it describes ${\mathcal S}_m$ in sufficient
detail by values of some quantities that can be ascribed to ${\mathcal S}$
alone independently of any observations. Second, it enables to calculate what
will be observed on ${\mathcal S}$ if some measurement is carried out. An
objective property of ${\mathcal S}$ is the value of a quantity of the first
kind. Thus, the values of observables of quantum system ${\mathcal S}$ are
{\em not} objective properties of ${\mathcal S}$. To ascribe values of
observables to ${\mathcal S}$ leads to contradictions such as contextuality
(for more detail, see \cite{hajicek6}). Here, we accept the old tenet: each
such value is only created during registration of ${\mathcal S}$ by some
apparatus ${\mathcal A}$ and it is an objective property of the composite
${\mathcal S} + {\mathcal A}$ rather than of ${\mathcal S}$ alone. It gives
only an indirect information about ${\mathcal S}$.

Fortunately, there are other observable properties of quantum systems. They
can be described by the following heuristic principle \cite{PHJT}.

{\bf Basic Ontological Hypothesis of QM} {\itshape A property is objective if
its value is uniquely determined by a preparation according to the rules of
standard QM. The "value" is the value of the mathematical expression
that describes the property and it may be more sophisticated than just a real
number. To observe an objective property, many registrations of one or more
observables are necessary.}

In fact, the Hypothesis just states explicitly the meaning that is tacitly
given to preparation by standard quantum mechanics. More discussion on the
meaning of preparation is in \cite{hajicek2,hajicek4}. In any case, prepared
properties can be ascribed to the prepared system without any problems or
paradoxes. Often, we shall use the Hypothesis as an heuristic principle: it
will just help to find some specific properties and then it will be forgotten,
that is, an independent assumption will be made that these properties can be
objective and each of them will be further studied.

It is easy to give examples of such objective properties. We can classify them
into {\em structural} and {\em dynamical}. The structural ones are eg mass,
charge, spin, structure of Hamiltonian of an isolated system, etc The
dynamical ones are eg state operator, average value and variance of an
observable, etc

Often, the Hypothesis meets one of the following two questions. First, how can
it be applied to cosmology, when there was nobody there at the Big Bang to
perform any state preparation? Second, a state preparation is an action of
some human subject; how can it result in an objective property? Both
objections result from the view that preparations are just some manipulations
in human laboratory. This is too narrow: even textbooks (see eg \cite{peres})
notice that preparations can also run spontaneously as natural processes, such
as the preparations of neutrinos that arrive at the Earth from the core of the
Sun. Moreover, the second objection is not much more than a pun. It is not
logically impossible that a human manipulation of a system results in an
objective property of the system. For example, pushing a snooker ball imparts
it a certain momentum and angular momentum that can be assumed to be objective
properties of the pushed ball.

\section{The space of quantum states} Quantum states of objects that we meet
in our everyday life are very different from wave functions. Thus, the usual
focus on wave functions is misleading if we want to understand classical
properties \cite{hajicek1}. We have to work with general states, which are
described by what is usually called "density matrices" \cite{peres}
or "state operators" \cite{BLM}.

\subsection{Mathematical preliminaries} Let us collect some mathematics on
quantum states that can be found in literature but is not generally known.

A bounded self-adjoint operator ${\mathsf a}$ is called {\em positive} if
$$
\langle \psi|{\mathsf a}|\psi \rangle \geq 0 \quad \forall |\psi \rangle \in
{\mathbf H}
$$
and {\em trace-class} if
$$
tr[\sqrt{{\mathsf a}^2}] < \infty\ .
$$
The left-hand side of the inequality can be shown to be a norm and the space
of trace-class operators completed with respect to this norm is a Banach
space. With the partial ordering ${\mathsf a} > {\mathsf b}$ defined by
operator ${\mathsf a} - {\mathsf b}$ being positive, it is an ordered Banach
space. (For more details, see \cite{ludwig1}, Appendix III and IV.)

What is the true state space of quantum mechanics?
\begin{df} Let ${\mathcal S}$ be a quantum system with Hilbert space ${\mathbf
H}$. Let ${\mathbf T}({\mathbf H})$ denote the ordered Banach space of
trace-class operators on ${\mathbf H}$. \underline{State space} ${\mathbf
T}({\mathbf H})^+_1$ of ${\mathcal S}$ is the set of all positive elements of
${\mathbf T}({\mathbf H})$ with trace 1.
\end{df} Hence, the space of states is the intersection of the unit sphere and
the positive cone of the Banach space ${\mathbf T}({\mathbf H})$. As the
positive cone has a non-trivial boundary in ${\mathbf T}({\mathbf H})$,
${\mathbf T}({\mathbf H})^+_1$ has a rich face structure.

The space ${\mathbf T}({\mathbf H})^+_1$ is not a linear space but it is {\em
convex}, that is, invariant with respect to {\em convex combination},
$$
w {\mathsf T}_1 + (1-w) {\mathsf T}_2 = {\mathsf T}\ ,
$$
where ${\mathsf T}_{1,2} \in {\mathbf T}({\mathbf H})^+_1$ and $w \in
[0,1]$. The states ${\mathsf T}_1$ and ${\mathsf T}_2$ are called {\em convex
components} of ${\mathsf T}$.
\begin{df}\label{facedf1} \underline{Face} ${\mathbf F}$ is a norm-closed
subset of ${\mathbf T}({\mathbf H})^+_1$ that is invariant with respect to
convex combinations and contain all convex components of any ${\mathbf T} \in
{\mathbf F}$.
\end{df} "Face" is an important notion of the mathematical theory of
convex sets.
\begin{thm}\label{facethm1} Every face ${\mathbf F} \subset {\mathbf
T}({\mathbf H})^+_1$ can be written as ${\mathbf F}({\mathsf T})$ for a
suitably chosen ${\mathsf T} \in {\mathbf F}({\mathsf T})$ where ${\mathbf
F}({\mathsf T})$ is the smallest face which contains ${\mathsf T}$.
\end{thm} For proof, see \cite{ludwig1}, p 76. Thus, each state operator lies
in some face. There is a useful relation between faces and projections:
\begin{thm}\label{facethm2} To each face ${\mathbf F}$ of ${\mathbf
T}({\mathbf H})^+_1$ there is a unique projection ${\mathsf P} : {\mathbf H}
\mapsto {\mathbf H}'$, where ${\mathbf H}'$ is a closed subspace of ${\mathbf
H}$, for which ${\mathsf T} \subset {\mathbf F}$ is equivalent to
$$
{\mathsf T} = {\mathsf P}{\mathsf T}{\mathsf P}\ .
$$
The map so defined between the set of faces and the set of projections is an
order isomorphism ie it is invertible and ${\mathsf P}_1 < {\mathsf P}_2$ is
equivalent to ${\mathbf F}_1 \subset {\mathbf F}_2$.
\end{thm} For proof, see \cite{ludwig1}, p 77. We shall denote the face that
corresponds to a projection ${\mathsf P}$ by ${\mathbf F}_{\mathsf P}$.

Clearly, intersection of two faces, if non-empty, is a face and a unitary map
of a face is a face. The next theorem shows that ${\mathbf F}({\mathsf T})$ is
not necessarily the set of all convex components of ${\mathsf T}$.
\begin{thm}\label{facethm3} Let ${\mathsf P}({\mathbf H})$ be
infinite-dimensional. Let ${\mathsf T}_1, {\mathsf T}_2 \in {\mathbf
F}_{\mathsf P}$ and be positive definite on ${\mathsf P}({\mathbf H})$. Then
$$
{\mathbf F}({\mathsf T}_1) = {\mathbf F}({\mathsf T}_2) = {\mathbf F}_{\mathsf
P}\ .
$$
Let $\{|k\rangle\}$ be an orthonormal basis of ${\mathsf P}({\mathbf H})$ and
let
\begin{equation}\label{suprem} \sup_k \frac{\langle k|{\mathsf
T}_1|k\rangle}{\langle k|{\mathsf T}_2|k\rangle} = \infty\ .
\end{equation} Then ${\mathsf T}_1$ is not a convex component of ${\mathsf
T}_2$.
\end{thm} For proof, see \cite{hajicek6}.
\begin{df}\label{facedf2} ${\mathsf T} \in {\mathbf T}({\mathbf H})^+_1$ is
called \underline{extremal} if it lies in a zero-dimensional face.
\end{df} For extremal states, we have:
\begin{thm}\label{facethm4} ${\mathsf T}$ is extremal iff ${\mathsf T} =
|\psi\rangle \langle \psi |$, where $\psi$ is a unit vector of ${\mathbf H}$.
\end{thm} For proof, see \cite{ludwig1}, p 78. Thus, extremal state can be
described by elements of ${\mathbf H}$, and only the $q$-representation of
such an element is a "wave function".

\subsection{Physical interpretation of the mathematics} Here, the mathematical
consequences of our heuristic principle will be worked out. The physical
meaning of elements of ${\mathbf T}({\mathbf H})^+_1$ is given by
\begin{assump} Let ${\mathcal S}$ be a quantum system with Hilbert space
${\mathbf H}$. Every ${\mathsf T} \in {\mathbf T}({\mathbf H})^+_1$ can be
prepared as a state of ${\mathcal S}$ and is then an objective property of
${\mathcal S}$.
\end{assump}

Next, we start to give a possible physical meaning to convex combinations.
\begin{df} Let ${\mathcal P}_1$ and ${\mathcal P}_2$ be two preparations of
${\mathcal S}$ and $w \in [0,1]$. \underline{Statistical} \underline{mixture},
\begin{equation}\label{statmixP} {\mathcal P} = \{(w,{\mathcal
P}_1),((1-w),{\mathcal P}_2)\}\ ,
\end{equation} of ${\mathcal P}_1$ and ${\mathcal P}_2$ is the following
preparation: Let ${\mathcal S}$ be prepared either by ${\mathcal P}_1$ or by
${\mathcal P}_2$ in a random way so that ${\mathcal P}_1$ is used with
probability $w$ and ${\mathcal P}_2$ with probability $1-w$.
\end{df}
\begin{assump} Let ${\mathcal P}_1$ and ${\mathcal P}_2$ be two preparations
of ${\mathcal S}$ and let they prepare states ${\mathsf T}_1$ and ${\mathsf
T}_2$. Then the statistical mixture (\ref{statmixP}) prepares state
$$
{\mathsf T} = w{\mathsf T}_1\ (+)_p\ (1-w){\mathsf T}_2\ .
$$
We call the right-hand side \underline{statistical decomposition} of ${\mathsf
T}$.
\end{assump} The purpose of sign "$(+)_p$" on the right-hand side is
to stress that this convex combination is a statistical decomposition. To be
aware of the distinction is very important for the understanding of the theory
of quantum measurement, see \cite{hajicek6}. For example, the theory of
quantum decoherence can achieve that the final state of the apparatus is a
convex combination of distinct pointer states but cannot conclude that it is a
statistical decomposition and must, therefore, resort to further assumptions
such as Everett interpretation \cite{schloss}. Let us stress that a
statistical decomposition of state ${\mathsf T}$ is not determined by the
mathematical structure of state operator ${\mathsf T}$ but by a preparation of
${\mathsf T}$.

Sometimes, one meets the objection that states $w{\mathsf T}_1\ (+)_p\
(1-w){\mathsf T}_2$ and $w{\mathsf T}_1 + (1-w){\mathsf T}_2$ of system
${\mathcal S}$ cannot be distinguished by any measurement. But this is only
true if the measurements are limited to registrations of observables of
${\mathcal S}$. If observables of arbitrary composite systems containing
${\mathcal S}$ are also admitted, then the difference between a statistical
decomposition and a convex combination can be found by measurements
\cite{survey}. This is exactly the argument against the decoherence theory
described in \cite{d'Espagnat}, p 171. Let us also emphasise that quantum
state statistics has nothing to do with the statistics of values of
observables.

As a mathematical operation, $(+)_p$ is commutative and associative. Thus, the
definitions and assumptions can be generalised to more than two preparations
and states. Moreover, state decomposition is invariant with respect to system
composition and unitary evolution \cite{hajicek2}.
\begin{df} Prepared state ${\mathsf T}$ of the form $w {\mathsf T}_1\ (+)_p\
(1-w) {\mathsf T}_2$ with $w \in (0,1)$ is called \underline{decomposable}.
\end{df} The meaning of a prepared state ${\mathsf T}$ being indecomposable is
that the ensemble ${\mathbf E}$ defined by repeating the preparation
${\mathcal P}$ of ${\mathsf T}$ has no sub-ensemble ${\mathbf E}'$ that can be
obtained by repeating preparation ${\mathcal P}'$ of state ${\mathsf T}' \neq
{\mathsf T}$.

Thus, to be decomposable or indecomposable are {\em physical} properties of
prepared states that are determined by the preparations of the states, not
just by {\em mathematical} existence of convex combinations. The faces just
restrict possible statistical decomposition of states. For example, any
extremal state is {\em indecomposable}. Decomposable states has been also
called "proper mixtures" \cite{d'Espagnat}, "direct
mixtures" \cite{ludwig1} or "gemenges" \cite{BLM,survey}.

There are prepared non-extremal states that are indecomposable. Consider
system ${\mathcal S}$ prepared in the EPR experiment (see eg \cite{peres}, p
150). ${\mathcal S}$ is a spin 0 system composite of two spin 1/2 systems
${\mathcal S}_1$ and ${\mathcal S}_2$. ${\mathcal S}$ is prepared in extremal
state $|\psi\rangle \langle\psi|$,
$$
|\psi\rangle = \frac{1}{\sqrt{2}}(|1+\rangle \otimes |2-\rangle -|1-\rangle
\otimes |2+\rangle)\ ,
$$
where $|1+\rangle$ is the state of the first particle with spin up etc, so
that the $z$ components of the spins of the two subsystems are
anti-correlated. Then the state of ${\mathcal S}_1$ is
\begin{equation}\label{epr1} tr_{{\mathcal S}_2}[{|\psi\rangle \langle\psi|}]
= \frac{1}{2}(|1+\rangle \langle 1+| + |1-\rangle \langle 1-|)\ ,
\end{equation} which is not extremal but the right-hand side on Eq.\
(\ref{epr1}) cannot be a statistical decomposition because this would imply
that Bell inequality were satisfied. Here, $tr_{{\mathcal S}_2}[{|\psi\rangle
\langle\psi|}]$ is a partial trace of $|\psi\rangle \langle\psi|$ over the
degrees of freedom of ${\mathcal S}_2$, see eg \cite{peres}.

By the way, the experiment can also be considered as a preparation of
${\mathcal S}_1$ in state (\ref{epr1}). This state is then an objective
property of ${\mathcal S}_1$ irrespectively of the entanglement of ${\mathcal
S}_1$ with ${\mathcal S}_2$. One can also say that $|\psi\rangle \langle\psi|$
is an entangled state,
$$
|\psi\rangle \langle\psi| \neq tr_{{\mathcal S}_2}[{|\psi\rangle
\langle\psi|}] \otimes tr_{{\mathcal S}_1}[{|\psi\rangle \langle\psi|}]\ ,
$$
but (\ref{epr1}) is not.

We observe that a quantum state is conceptually very different from a state in
Newtonian mechanics. It may be helpful to look at some important
differences. Let us define a state of a Newtonian system as point $p$ of the
phase space of the system, ${\mathbf \Gamma}$. Newtonian state defined in this
way is generally assumed to satisfy:
\begin{enumerate}
\item {\em Objectivity}: a state is an objective property.
\item {\em Universality}: any system is always in some state.
\item {\em Exclusivity}: a system cannot be in two different states
simultaneously.
\item {\em Completeness}: any state of an isolated system represents the
maximum information that can exist about this system.
\item {\em Locality}: the state of a systems determines the position of the
system.
\end{enumerate} An incomplete information about the state of a system can be
described by a probability distributions on ${\mathbf \Gamma}$. Such a
distribution is sometimes called {\em statistical state}. In any case, we
distinguish a state from a statistical state.

A quantum state is an element of ${\mathbf T}({\mathbf H})^+_1$ and the
comparison to Newtonian states is described by:
\begin{enumerate}
\item There is objectivity: a prepared state is an objective property.
\item There is no universality: a system need not be in any state (an example
is a particle ${\mathcal S}$ in a system ${\mathcal S}'$ of identical
particles and we assume that a state of ${\mathcal S}'$ has been prepared but
that of ${\mathcal S}$ has not \cite{hajicek4}).
\item There is no exclusivity: a system can be in several states
simultaneously (such as ${\mathsf T}_1$ and $w {\mathsf T}_1\ (+)_p\ (1-w)
{\mathsf T}_2$ above).
\item There is no completeness: a state operator alone does not contain any
information on the statistical decomposition of the prepared state. However,
if an indecomposable state of an isolated system is given, no more knowledge
on the system can objectively exist than that given by the state.
\item Quantum states are non-local: most states of a single particle do not
determine its position, but two detectors at different positions will give
anti-correlated results (\cite{hajicek4}).
\end{enumerate} In particular, we have the following correspondences:
$$
\begin{array}{lll} \text{quantum indecomposable state} & \longleftrightarrow &
\text{Newtonian state} \\ \text{quantum decomposable state} &
\longleftrightarrow & \text{Newtonian statistical state}
\end{array}
$$
There is, of course, the difference that a decomposable state is described by
an element of ${\mathbf T}({\mathbf H})^+_1$, whereas a Newtonian statistical
state is not an element of ${\mathbf \Gamma}$. From the point of view of both
statistics of states and statistics of values of observables, textbooks focus
on the difference between {\em pure} (extremal) and {\em mixed} (non-extremal)
states is rather misleading.

An important consequence of our interpretation is the following. We consider
any indecomposable state as a complete description of the reality of the
system. Thus, the collapse of wave function and analogous processes must be
viewed as physical processes, not just as changes of our information about the
system \cite{hajicek2,hajicek4,hajicek6}.

\subsection{Simple objective properties} In quantum mechanics, simple
objective properties can be defined as follows.
\begin{df} Let $f : {\mathbf T}({\mathbf H})^+_1 \mapsto {\mathbb R}$. A
\underline{simple objective property} is defined by proposition "$f$ has
value $a$".
\end{df} As states are uniquely determined by preparations, and values of $f$
by the states, the simple properties are objective. Each simple property is
equivalent to a subset of ${\mathbf T}({\mathbf H})^+_1$ described by
$$
\{ {\mathsf T} \in {\mathbf T}({\mathbf H})^+_1 \ |\ f({\mathsf T}) = a \}\ .
$$
Quantum simple properties form a Boolean lattice with respect to logical
operations of union and intersection that is isomorph to the lattice of
subsets of ${\mathbf T}({\mathbf H})^+_1$. This is clearly different from the
so-called "quantum logic", which holds for properties defined by
propositions "Projection ${\mathsf P}$ has value $\eta$" with $\eta
= 0,\ 1$ (see eg \cite{bub}). The reason is that our simple properties are
defined by preparations while those of quantum logic by registrations because
${\mathsf P}$ is an observable and $\eta$ its value.

Quantum simple properties are analogous to Newtonian properties that can be
defined as real functions on ${\mathbf \Gamma}$. For example, a system having
energy $E$ is a property equivalent to the subset of points of ${\mathbf
\Gamma}$ at which the energy has value $E$. Newtonian properties also form a
Boolean lattice.

A very important example of quantum simple property is average $tr[{\mathsf A}
{\mathsf T}]$ of observable ${\mathsf A}$ in state ${\mathsf T}$. One could
ask how comes that average of ${\mathsf A}$ is an objective property while its
values are not? But this is simple: the average can be considered as a
predetermined condition on the values that are to be registered: in large
numbers these values must add to the average. Thus, the average is an example
of a property that can be measured by a large number of individual
registrations.

Clearly there are enough simple objective properties to describe the dynamical
situation of any quantum object completely. Thus, our theoretical models of
real objects can ascribe them enough objective properties, as required. More
discussion is in \cite{PHJT,survey,hajicek6}.

\section{Conclusion and outlook}
\begin{itemize}
\item Constructive Realism by Giere---a kind of relatively strong realism---is
shown to be compatible with quantum mechanics.
\item The resulting realist interpretation is indispensable for our method to
construct quantum models of classical properties \cite{hajicek1,hajicek6} and
of measurement processes \cite{hajicek2,hajicek3,hajicek6}.
\end{itemize}

\subsection*{Acknowledgements} The author is indebted to Petr Jizba, Karel
Kucha\v{r} and Ji\v{r}\'{\i} Tolar for discussion.

\end{document}